\documentclass[11pt,twoside,a4paper]{article}
\usepackage{amsmath}
\usepackage{graphicx}
\usepackage{caption}
\usepackage{subcaption}
\usepackage[numbers,sort&compress,square]{natbib}

\begin{document}

\title{Semiclassical approach to atomic decoherence by gravitational waves}
\author{Diego A. Qui\~nones, Benjamin Varcoe}
\maketitle

\begin{abstract}
A new heuristic model of interaction of an atomic system with a gravitational wave is proposed. In it, the gravitational wave alters the local electromagnetic field of the atomic nucleus, as perceived by the electron, changing the state of the system. The spectral decomposition of the wave function is calculated, from which the energy is obtained. The results suggest a shift in the difference of the atomic energy levels, which will induce a small detuning to a resonant transition. The detuning increases with the quantum numbers of the levels, making the effect more prominent for Rydberg states. We performed calculations on the Rabi oscillations of atomic transitions, estimating how they would vary as a result of the proposed effect.
\end{abstract}

\section{Introduction}
In general relativity, gravity is the curvature of the space-time continuum produced by the mass of objects \cite{schwarzschild16}. Similar to how accelerating electrical charges produce electromagnetic waves, accelerating masses will produce ripples in the fabric of space-time \cite{einstein16,einstein18}, which are called gravitational waves (GWs). They are predicted to exist in a very wide range of frequencies, depending on the source that generate them \cite{poisson93,haehnelt94,ryan95,ryan97,turner97,hogan98,hogan00,damour01,benacquista02,durrer10}, but even the most energetic ones (like the ones produced by rotating neutron stars \cite{bildsten98}) will only produce a very small distortion of space, making them very hard to detect. \\
Although GWs have been previously observed indirectly \cite{taylor94}, it was only recently that a direct detection was performed by analyzing the signal of very big interferometers \cite{abbot16,abbot17,abbot18}. These experiments not only open the window to a new way to observe the universe but also showed that gravitational waves have physical effects that can be detected on Earth using currently available technology. \\
Some theories have calculated the effect that gravitational waves have on particles, where decoherence is expected to arise \cite{blencowe13,oniga16}. Our approach is to obtain the change in the energy levels of an hydrogen-like atom \cite{zhao07,parker82} and analyze how the properties of the atom can be altered in order to make the shift significant enough to appear in the spectroscopic signal of said atoms. \\ 
In our model, a gravitational wave passing by an atom will curve the space, deforming the electromagnetic potential from the nucleus as felt by the electron, which will change the energy of the system. The new energy is derived by estimating the wave function of the electron after the interaction.
The results of these calculations suggest that the atom will suffer a small shift in the energy of its transitions, being this more prominent for high-energy levels. By maximizing the order of the perturbation, it could be possible to apply our model for the detection and characterization of gravitational waves. Because the only constrain we impose to the model is for the interacting gravitational wave to have a wavelength considerably bigger than the size of the atom, the proposed detection scheme could be applied to the detection of relic gravitational waves and the high-frequency range of the stochastic gravitational wave background \cite{ligo09,lasky16,wang16,grishchuk05,copeland09,andriot17}, which current experiments cannot observe, providing a powerful tool for the study of the universe.

\section{Wave-Function shift}
\begin{figure}
  \centering
    \includegraphics[width=1 \textwidth]{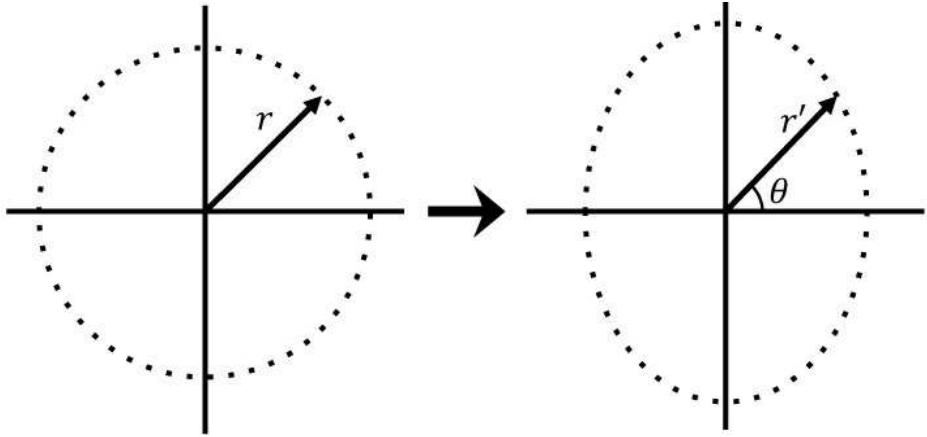}
    \caption{Representation of the space strain resulting from an incoming gravitational wave, in the plane transversal to the direction of propagation.} \label{fig1}
\end{figure}

Let us consider an hydrogen-like atom in an excited state. The potential felt by the valence electron will be given by
\begin{equation}
V_n = - \frac{k_B  { \ } e^2}{r_n} { \ },
\end{equation}
where $k_B$ is the Boltzmann constant, $e$ is the elementary charge and $r_n$ is the distance that depends on the principal quantum number $n$ and contains the correction due to the screening of the nucleus (quantum defect) \cite{hezel92}. 
An incoming gravitational wave will compress the space along one of the axis of the plane transversal to its propagation direction and expands the space along the perpendicular direction within the plane (see Fig. \ref{fig1}), producing a change in the coordinates 
\begin{equation} \label{vecr}
\vec{r} \to \vec{r}'(r,\theta,\phi) { \ } .
\end{equation}
This will result in the electron perceiving the nucleus at a different distance, altering the effective potential,
\begin{equation} 
V_n(\vec{r}) \to V'(\vec{r}') { \ } .
\end{equation}
To better understand the potential shift, picture the GW modifying the wavelength of the force-carrying photons between the electron and nucleus: Their wavelength will be compress along a certain angular direction and expanded along the perpendicular angle, changing accordingly the momentum exchange between the charges.  \\
The potential is expected to change smoothly, as the GW is a gradual distortion of space (adiabatic approximation). This assumption should specially hold for GWs with wavelength bigger than the size of the atom, which is typically the case.
The adiabatic change will make the wave function of the atom to evolve into the appropriate eigenstate of the new potential 
\begin{equation}
\psi(V_n) \to \psi' = \psi' (V') { \ } .
\end{equation}
Rather than finding the eigenstates related the new perceived potential, we estimate the wave function as evolving into the distorted space coordinates,
\begin{equation} \label{psivrtpspepsvrtvrp}
\psi(\vec{r}) \to \psi' = \psi (\vec{r} \to \vec{r}') { \ } .
\end{equation}
This transformation gives back the original wave function in the absence of the perturbation ($\vec{r}' = \vec{r}$), which is required for the adiabatic approximation. In the classical picture, Eq. (\ref{psivrtpspepsvrtvrp}) can be interpreted as the stable orbital becoming an ellipse as result of the loss of radial symmetry in the electric potential in order to valance it with the kinetic energy, which is roughly the meaning of the Schr\"{o}dinger equation. \\
In order to obtain the energy of the system, we calculate the resulting wave function $\psi'$ in terms of the eigenstates of the unperturbed potential,
\begin{equation} \label{psipnlmcnlmprtp}
{\psi'} = \sum_{n, l, m} C_{n, l, m} \psi_{n, l, m} (r, \theta, \phi) { \ },
\end{equation}
where $n$, $l$ and $m$ are the quantum numbers. We are not suggesting that the wave function is projected into the eigenstates (as this will imply an non-adiabatic process), but just using an equivalent representation of the transformed wave function in the original base. 
 The coefficients $C_{n, l, m}$ in Eq. \ref{psipnlmcnlmprtp}) can be calculated with
\begin{equation}\label{cnlm}
C_{n, l, m} = \int_0^{\infty} r^2 dr \int_0^{\pi} sin\theta { \ } d\theta { \ } \int_0^{2 \pi} d\phi { \ } \psi_{n, l, m}^{*} {\psi'} { \ } .
\end{equation}
Solving numerically Eq. (\ref{cnlm}) for an arbritrary change  requires a lot of computational time, even for the most simple atomic system. 
To simplify calculations, we use the following approximation:
First consider an hydrogen-like atom in an excited state with principal quantum number $n_0$, and no angular momentum (azimuthal quantum number $l = 0$). The initial wave function of the system is
\begin{equation} \label{psi0}
\psi_{n_0,0} = \sqrt{\left( \frac{2}{n_0 a_0} \right)^3 \frac{(n_0 -1)!}{8 \pi n_0 (n_0!)^3} } e^{-\frac{r}{n_0 a_0}} L^1_{n_0
- 1} \left( \frac{2r}{n_0 a_0} \right) { \ } ,
\end{equation} 
where $a_0$ is the Bohr radius and $L^1_{n_0- 1}$ is the corresponding associated Laguerre polynomial \cite{griffiths}. 
In the plane transversal to the GW's direction, we model the distortion of the space as 
\begin{equation} \label{rprime}
r \to r'= r \frac{1-S_p}{\sqrt{cos^2\theta + \left( \frac{1-S_p}{1+S_p} \right)^2 sin^2 \theta}} \equiv r A_{\theta} { \ } ,
\end{equation} 
where $S_p$ represents the strain on space in said plane. 
For GWs with wavelength significantly bigger than the expected size of the atom, we can consider the strain $S_p$ a constant. This allows us to substitute Eq. (\ref{rprime}) directly into Eq. (\ref{psi0}), such that the wave function is transformed as
\begin{equation}\label{psip}
{\psi'}  = \sqrt{\left( \frac{2}{n_0 a_0} \right)^3 \frac{(n_0 -1)!}{8 \pi n_0 (n_0!)^3} } e^{-\frac{r A_{\theta}}{n_0 a_0}} L^1_{n_0- 1} \left( \frac{2 r A_{\theta}}{n_0 a_0} \right) { \ } .
\end{equation}    
In order to find an analytical solution to Eq. (\ref{cnlm}), we use the mathematical identity
\begin{equation} \label{l1sl}
 L^1_{n_0- 1} \left( \frac{2 r A_{\theta}}{n_0 a_0} \right) =  e^{-\frac{2 r}{n_0 a_0}(1-A_{\theta})  } \sum_{k=0} \frac{(1-A_{\theta})^k}{k!} \left( \frac{2 r}{n_0 a_0} \right)^k  L^{1+k}_{n_0- 1} \left( \frac{2 r }{n_0 a_0} \right),
\end{equation}
and the approximations 
\begin{equation}
e^{- \frac{r}{n_0 a_0} (3 - 2 A_{\theta})} \approx e^{- \frac{r}{n_0 a_0}} { \ } , 
\end{equation}
\begin{equation} \label{1mat}
(1 - A_{\theta})^k \approx \left[ S_p cos(2 \theta) \right]^k { \ } ,
\end{equation}
which are valid for very small values of the strain constant ($S_p \ll 1$)\cite{abbot16}. Using Eq. (\ref{l1sl}) to Eq. (\ref{1mat}) with the identity in Eq. (\ref{psip}) allows us to separate the wave function after the interaction into a radial and an angular part 
\begin{equation} \label{psisum}
{\psi'} = \sum_{k=0} \frac{{S_p}^k}{k!} {R'}_{n_0,k} (r) { \ } {Y'}_{k} (\theta) { \ } ,
\end{equation}
\begin{equation}
{R'}_{n_0,k} (r) = \sqrt{\left( \frac{2}{n_0 a_0} \right)^3 \frac{(n_0 -1)!}{2 n_0 (n_0 !)^3} } e^{- \frac{r}{n_0 a_0}}  \left( \frac{2 r}{n_0 a_0} \right)^k L^{1+k}_{n_0- 1} \left( \frac{2 r }{n_0 a_0} \right)  { \ } , 
\end{equation}
\begin{equation}
{Y'}_{k} (\theta) = \frac{1}{\sqrt{4 \pi}} cos^k (2 \theta) L_{n_0 - 1}^{1 + k} \left( \frac{2 r}{n_0 a_0} \right)  { \ } ,
\end{equation}
which is similar to the solution of the Schr\"{o}dinger equation for hydrogen-like atoms ($ \psi_{n, l, m} = R_{n,l} (r) { \ } Y_l^m (\theta, \phi)$). 
The integral of the radial component is equal to
\begin{equation} \label{intrn}
\int_0^{\infty}{R'}_{n_0,k} (r) { \ } R_{n,l} (r) { \ } dr = \sqrt{\frac{(n_0 -1)! (n_0 - l - 1)!}{[n_0 ! (n_0 + l)!]^3}} \frac{[(n_0 + k)!]^3}{(n_0 - l - 1)!} \delta_{n_0 , n}{ \ },
\end{equation}
where $\delta_{n_0 , n}$ is the Kronecker delta.
Solving the integral for the angular component yields to
\begin{equation} \label{inttf}
\int_0^{\pi}  \int_0^{2 \pi} sin\theta { \ } {Y'}_{k} (\theta) { \ } Y_l^m (\theta, \phi) { \ } d\phi { \ } d\theta = 
  \begin{cases}
   \sqrt{2 l + 1} { \ }\Theta_{k,l} &\text{for } l = 0,2,4,... \\
    0&\text{for } l = 1,3,5,...\\
  \end{cases}
\end{equation}
This equation indicates that after the gravitational interaction, the atom will be perceived as a superposition of states with even azimuthal number. Some of the non-trivial values of $\Theta_{k,l}$ are shown in Table (\ref{tfkl}). 
\begin{table}
\centering
\caption{Value of the component $\Theta_{k,l}$ by solving Eq. (\ref{inttf}) for different values of $k$ and $l$.}
\label{tfkl}
\begin{tabular}{l|c|c|c|c|}
\cline{2-5}
& \textbf{\textit{l}=0} & \textbf{\textit{l}=2} & \textbf{\textit{l}=4} & \textbf{\textit{l}=6} \\ \hline
\multicolumn{1}{|l|}{\textbf{\textit{k}=0}} & 1    & 0     & 0    & 0   \\ \hline
\multicolumn{1}{|l|}{\textbf{\textit{k}=1}} & -1/3 & 4/15 & 0 & 0   \\ \hline
\multicolumn{1}{|l|}{\textbf{\textit{k}=2}} &  7/15 & -8/105 & 32/315 & 0  \\ \hline
\multicolumn{1}{|l|}{\textbf{\textit{k}=3}} &  -9/15 & 4/21 & -32/1155 & 128/3003 \\ \hline
\end{tabular}
\end{table}
With the product of Eq. (\ref{intrn}) and Eq. (\ref{inttf}) we obtain the coefficients $C_{n,l,m}$ of Eq. (\ref{psisum}). Because the terms of the summation are proportional to ${S_p}^k$, which is expected to be extremely small ($\sim 10^{-20}$) \cite{blencowe13}, we disregard all the terms in Eq. (\ref{psisum}) for $k \geq 2$, obtaining
\begin{equation} \label{psisum2}
{\psi'}_{n_0,0,0} \approx C_{0} (n_0) \psi_{n_0,0,0} + C_{2} (n_0) \psi_{n_0,2,0} { \ } ,
\end{equation}
with
\begin{eqnarray} \label{c0c2}
&C_{0} (n_0) = 1- \frac{S_p}{3} (n_0 + 1)^3  { \ }, \\
&C_{2} (n_0) =  S_p \frac{4 (n_0 +1)}{3 (n_0 +2 )^2} \sqrt{({n_0}^2-1) ({n_0}^2-4)/5}  { \ }.
\end{eqnarray}
The coefficient $C_0$ will be several orders of magnitude bigger than $C_2$, meaning the change in the system is expected to be very small.
Eq. (\ref{c0c2}) can be interpreted as the GW coupling the energy levels associated with $\psi_{n_0,0,0}$ and $\psi_{n_0,2,0}$, similar to the expected quadrupole interaction \cite{einstein18}. A transition between these energy levels can be attributed to an interaction with the graviton (with $l=2$), which will occur with probability $|C_2|^2$. \\
For the general case, following the same calculations (Eq. (\ref{cnlm}) to Eq. (\ref{1mat})) for an atom in an initial state with principal quantum number $n_0 \geq 2$ and azimuthal quantum number $l_0 \geq  1$ give us that the wave functions evolves into an state that can be expressed as a superposition of eigenfunctions with azimuthal number equal to the initial one plus or minus factors of two ($l= l_0 \pm 2, 4, 6,...$).
With similar arguments to the case of initial angular momentum $l_0 = 0$, the resulting wave function can be approximated to 
\begin{equation} \label{pspc}
{\psi'} \approx C_0 \psi_{n_0, l_0} + C_{+2} \psi_{n_0, l_0 +2} + C_{-2} \psi_{n_0, l_0 - 2} { \ },
\end{equation}
where the coefficients $C_i$ are given by
\begin{equation} \label{c2pm}
\begin{split}
&C_0 = 1 - S_p \frac{(n_0 + l_0 +1)^3}{(2 l_0 - 1)(2 l_0 + 3)} \\
&C_{+2} =  2 S_p \frac{(l_0 + 1)(l_0 + 2)}{2 l_0 + 3} \sqrt{\left( \frac{n_0 + l_0 +1}{n_0 + l_0 + 2} \right)^3 \frac{(n_0 - l_0 -1)(n_0 - l_0 - 2)}{(2 l_0 +1)(2 l_0 +5)} } \\
&C_{-2} =2 S_p \frac{l_0 (l_0 -1) (n_0 + l_0 +1)^3}{2 l_0 -1} \sqrt{ \frac{(n_0 + l_0)^3 (n_0 + l_0 -1)^3}{(n_0 - l_0)(n_0 - l_0 + 1)(2 l_0 +1)(2 l_0 -3)} }
\end{split}
\end{equation}
Because the coefficients are non-linear functions of the quantum numbers, each energy level will be displaced by a different degree. This will result in a small shift in the energy of the atomic transitions.   

\section{Transition detuning}
\begin{figure}
 \centering
    \includegraphics[width=.7 \textwidth]{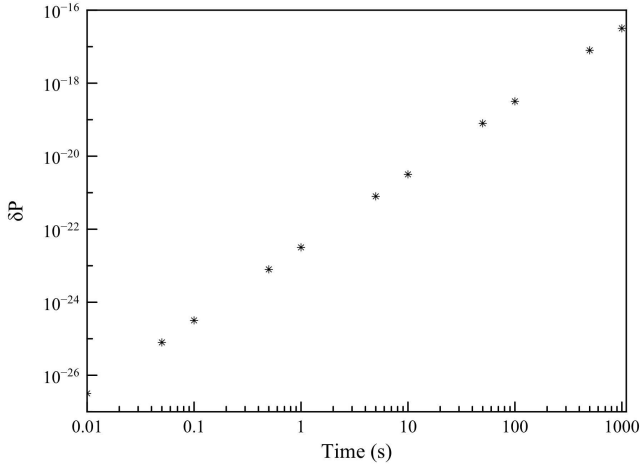}
    \caption{Deviation in the Rabi cycle for a 50S - 51P transition with a Rabi frequency of 47 kHz \cite{brune96}, which arises from the interaction with a gravitational wave with peak strain of $Sp = 10^{-20}$.} \label{fig2}
\end{figure}

An atom interacting with a gravitational wave experiences a change in its energy levels
\begin{equation}
E_{n,l} \to E_{n,l} ' { \ },
\end{equation} 
which depends on the coefficients $C_0$, $C_{+2}$ and $C_{-2}$,
\begin{equation} \label{epnl}
E_{n,l} ' = {C_0}^2 E_{n,l} + {C_{+2}}^2 E_{n,l+2} + {C_{-2}}^2 E_{n,l-2} { \ }.
\end{equation} 
Therefore, the difference in energy between two distinct energy levels,
\begin{equation}
\Delta E = E_{2} - E_{1} { \ },
\end{equation}
is shifted by a factor $\delta$,
 \begin{equation}
\Delta E' = E_{2}' - E_{1}'  = \Delta E + \delta { \ },
\end{equation}
which will be the detuning for light used to drive a transition between states with energy  $E_1$ and $E_2$.
From Eq. (\ref{epnl}) we have that
 \begin{equation}
 E_i ' - E_i = - E_i \left( 1 - {C_{0}}^2 - {C_{+2}}^2 \frac{E_{i +2}}{E_i} -  {C_{-2}}^2 \frac{E_{i -2}}{E_i} \right) { \ }.
\end{equation}
Using the values of $C_j$ indicated in Eq. (\ref{c2pm}) yields
 \begin{equation}
 E_i ' - E_i = - E_i \left[ 2 S_p \frac{(n_i + l_i + 1 )^3}{(2 l_i - 1)(2 l_i + 3)} + \kappa ({S_p}^2) \right]{ \ },
\end{equation}
where $\kappa$ is a factor proportional to ${S_p}^2$. Because this term is very small, we can approximate
 \begin{equation}
 E_i ' - E_i \approx - 2 S_p \frac{ E_i  (n_i + l_i + 1 )^3}{(2 l_i - 1)(2 l_i + 3)} { \ }.
\end{equation}
With this, we finally arrive to the relation
 \begin{equation}\label{dsp}
\delta = - 2 S_p  \left[ \frac{ E_2  (n_2 + l_2 + 1 )^3}{(2 l_2 - 1)(2 l_2 + 3)} - \frac{ E_1  (n_1 + l_1 + 1 )^3}{(2 l_1 - 1)(2 l_1 + 3)}  \right]{ \ }.
\end{equation}
Light with energy $\Delta E$ will be therefore detuned by $\delta \sim  S_p \Delta E$ to the atom's transition. Eq.(\ref{dsp}) indicates that the detuning can be increased by using transitions between states with high quantum numbers \emph{i.e.} Rydberg states, which will be easier to detect \cite{quinones16}. For states with  $n \sim 50$ \cite{singer04,kubler10}, the detuning $\delta$ can increase by a factor of $10^5$, even for transitions of states with close quantum numbers. 
Using Rydberg atoms for gravitational wave detection has been proposed by previous studies \cite{fischert94,pinto95,quinones17}, further supporting our results. \\
The experimental observation of the proposed effect could be very difficult, even using the highly excited states previously mentioned. The analysis of the spectrum of extraplanetary Rydberg atoms could provide an advantage given the extremely high energies at which they can be found \cite{gnedin09,brooks01}. As an example, the H110$\alpha$ emission of the Carina nebula (4.8 GHz)\cite{brooks01} would experience a change in its wavelength of $5.6 \times 10^{-16}$m, which could be measured using extremely accurate interferometry, like the one currently applied in gravitation wave detectors \cite{abbot16,abbot17}.  Other studies have suggested a shift in the spectrum for excited atoms in regions with high spacetime curvature, but to achieve a shift of the order of the previous example it would require a characteristic radius of curvature of  $\sim 100$ Km,  involving being in the vicinity of a superdense object such a neutron star \cite{parker82}.  \\
The detuning could also have an observable effect in Rabi oscillations  \cite{johnson08} for high energy states. For a two-level system, the probability of finding it in the excited state will evolve as
\begin{equation}\label{pedt}
P_e (\Delta,t) = \frac{\omega^2}{\Delta^2 + \omega^2} \sin^2 \left( \frac{\sqrt{\Delta^2 + \omega^2}}{2} t \right) { \ } ,
\end{equation}
where $\omega$ is the Rabi frequency and $\Delta$ is the frequency detuning. 
The Rabi cycle will deviate from its resonant dynamics $P_e (\Delta = 0,t)$ by
\begin{equation}\label{dpt}
\delta P = P_e (0,t) - P_e (\Delta,t) =  \sin^2 \left( \frac{\omega}{2} t \right) - \frac{\omega^2}{\Delta^2 + \omega^2} \sin^2 \left( \frac{\sqrt{\Delta^2 + \omega^2}}{2} t \right) 
\end{equation}
When the detuning is much smaller than the Rabi frequency ($\Delta \ll \omega$), we can use the approximation
\begin{equation} 
\sqrt{\Delta^2 + \omega^2} \approx \omega + \Delta^2/2\omega { \ }, 
\end{equation}
which can be substituted in Eq. (\ref{dpt}) to obtain the approximation 
\begin{equation}\label{dpa}
\delta P \approx \sin^2 \left( \frac{\omega}{2} t \right) \left[1- \frac{\omega^2}{\Delta^2 + \omega^2}  \cos^2 \left( \frac{\Delta^2}{4 \omega} t \right) \right] { \ }.
\end{equation}
Using Eq. (\ref{dpa}) for short times (compared to the period of the gravitational wave), we get that the deviation can be approximated as
\begin{equation}
\delta P \approx \left(\frac{\Delta^2}{4 \omega} t \right)^2 { \ }.
\end{equation}
This equation indicates that the deviation increases with the detuning and decreases with the Rabi frequency, making again the effect more prominent in transitions of states with high quantum numbers \cite{dutta01}. For comparison, the deviation in the Rabi cycle $\delta P$ for the transition 50S - 51P of a Rubidium atom \cite{brune96} will be $10^4$ higher than the one for the transition 1S - 2P in the same atom. \\
For longer times ($t >  \pi \omega/\Delta^2$), the deviation will be mostly due to the increased frequency of the cycle. In this case, according to Eq. (\ref{dsp}), the ideal transition will be obtained by maximizing the energy difference between the levels \cite{dudin12}.  \\
If a system undergoing Rabi oscillations is measured at times corresponding to N completed cycles ($t =2N\pi/\omega$), the expected deviation will be
\begin{equation}
\delta P (t=2N\pi/\omega) \approx  \left( \frac{N \pi \Delta^2}{2 \omega^2} \right)^2 { \ }.
\end{equation}
In Fig. (\ref{fig2}) we show the expected measured deviation for completed cycles using the mentioned 50S - 51P transition.
In this figure it can be seen that the effect becomes significant for very long times, which may require experiments to implement a high-Q cavity in order to extend the coherence time of the atoms by suppressing most of the undesired transitions \cite{walther06,varcoe00}; this could also help to inhibit the interaction of  surrounding electromagnetic radiation, which otherwise may conceal the proposed effect. For the excitation levels used in our example, a high-Q cavity has been measured to increase the coherence times of the cavity-atom system to $0.02$ s \cite{deleglise08}, limiting the capability of observing the proposed effect by the achievable resolution of the state population within this timeframe. 
Although a big ensemble  of Rydberg atoms can be prepared to improve the probability of detection \cite{dudin12}, collective effects should be taken into consideration \cite{quinones17} especially for such highly excited atoms, as they exhibit long range correlations \cite{browaeys13,lukin01,lee12}.

\section{Conclusion}
We have modeled the effect of a gravitational wave passing through an atom as a distortion of the electromagnetic potential perceived by the electron. We calculated the resulting wave function as a superposition of the initial state and eigenstates with azimuthal quantum number that differ by two from the initial one. Our calculations indicate that the energy of the transitions in the atom will be shifted as a result of the interaction. Light with frequency equal to a atomic transition will then become slightly off-resonance, inducing a small deviation in the Rabi cycle of the atom. The deviation will increase drastically for transitions involving Rydberg states, which can make the proposed effect easier to detect. The proposed effect presents the possibility of using atomic spectroscopy for detection of gravitational waves. Because the model only assumes the gravitational wave as having a wavelength bigger than the atomic scale, observations could be done in frequency ranges different from those of large scale interferometry detectors. The capability of the experiments to distinguish between the different gravitational wave frequencies, and therefore their sources, will depend on the state shift resolution within the period of the analyzed gravitational waves. 

\bibliographystyle{unsrt}
\bibliography{aps_main}

\providecommand{\noopsort}[1]{}\providecommand{\singleletter}[1]{#1}%
\begin{thebibliography}{48}
\providecommand{\natexlab}[1]{#1}
\providecommand{\url}[1]{\texttt{#1}}
\expandafter\ifx\csname urlstyle\endcsname\relax
  \providecommand{\doi}[1]{doi: #1}\else
  \providecommand{\doi}{doi: \begingroup \urlstyle{rm}\Url}\fi

\bibitem[Andriot and Gomez(2017)]{andriot17}
D.~Andriot and G.~Lucena Gomez.
\newblock Signatures of extra dimensions in gravitational waves.
\newblock \emph{Journal of Cosmology and Astroparticle Physics}, 2017:\penalty0
  048, 2017.

\bibitem[Benacquista(2002)]{benacquista02}
M.~J. Benacquista.
\newblock Gravitational radiation from black hole binaries in globular
  clusters.
\newblock \emph{Classical and Quantum Gravity}, 19:\penalty0 1297, 2002.

\bibitem[Bildsten(1998)]{bildsten98}
L.~Bildsten.
\newblock Gravitational radiation and rotation of accreting neutron stars.
\newblock \emph{The Astrophysical Journal Letters}, 501:\penalty0 L89, 1998.

\bibitem[Blencowe(2013)]{blencowe13}
M.~P. Blencowe.
\newblock Effective field theory approach to gravitationally induced
  decoherence.
\newblock \emph{Phys. Rev. Lett.}, 111:\penalty0 021302, 2013.

\bibitem[Brooks et~al.(2001)Brooks, Storey, and Whiteoak]{brooks01}
K.~J. Brooks, J.~W.~V. Storey, and J.~B. Whiteoak.
\newblock H110a recombination-line emission and 4.8-ghz continuum emission in
  the carina nebula.
\newblock \emph{Monthly Notices of the Royal Astronomical Society},
  327:\penalty0 46, 2001.

\bibitem[Browaeys and Lahaye(2013)]{browaeys13}
A.~Browaeys and T.~Lahaye.
\newblock Interacting cold rydberg atoms: a toy many-body system.
\newblock \emph{Seminaire Poincare}, 17:\penalty0 125, 2013.

\bibitem[Brune et~al.(1996)Brune, Schmidt-Kaler, Maali, Dreyer, Hagley,
  Raimond, and Haroche]{brune96}
M.~Brune, F.~Schmidt-Kaler, A.~Maali, J.~Dreyer, E.~Hagley, J.~M. Raimond, and
  S.~Haroche.
\newblock Quantum rabi oscillation: A direct test of field quantization in a
  cavity.
\newblock \emph{Phys. Rev. Lett.}, 76:\penalty0 1800--1803, 1996.

\bibitem[Collaboration and Collaboration(2009)]{ligo09}
The LIGO~Scientific Collaboration and The~Virgo Collaboration.
\newblock An upper limit on the stochastic gravitational-wave background of
  cosmological origin.
\newblock \emph{Nature}, 460:\penalty0 990--994, 2009.

\bibitem[Copeland et~al.(2009)Copeland, Mulryne, Nunes, and Shaeri]{copeland09}
E.~J. Copeland, D.~J. Mulryne, N.~J. Nunes, and M.~Shaeri.
\newblock Gravitational wave background from superinflation in loop quantum
  cosmology.
\newblock \emph{Phys. Rev. D}, 79:\penalty0 023508, 2009.

\bibitem[D.~A.~Quinones et~al.()D.~A.~Quinones, Varcoe, and Wang]{quinones17}
T.~Oniga D.~A.~Quinones, B.~T.~H. Varcoe, and C.~H.-T. Wang.
\newblock Quantum principle of sensing gravitational waves: From the zero-point
  fluctuations to the cosmological stochastic background of spacetime.
\newblock Phys. Rev. D (unpublished, accepted 20 July 2017).

\bibitem[Damour and Vilenkin(2001)]{damour01}
T.~Damour and A.~Vilenkin.
\newblock Gravitational wave bursts from cusps and kinks on cosmic strings.
\newblock \emph{Phys. Rev. D}, 64:\penalty0 064008, 2001.

\bibitem[Deleglise et~al.(2008)Deleglise, Dotsenko, Sayrin, Bernu, Brune,
  Raimond, and Haroche]{deleglise08}
S.~Deleglise, I.~Dotsenko, C.~Sayrin, J.~Bernu, M.~Brune, J.~M. Raimond, and
  S.~Haroche.
\newblock Reconstruction of non-classical cavity field states with snapshots of
  their decoherence.
\newblock \emph{Nature}, 455:\penalty0 510--514, 2008.

\bibitem[Dudin et~al.(2012)Dudin, Li, Bariani, and Kuzmich]{dudin12}
Y.~O. Dudin, L.~Li, F.~Bariani, and A.~Kuzmich.
\newblock Observation of coherent many-body rabi oscillations.
\newblock \emph{Nat. Phys.}, 8:\penalty0 790--794, 2012.

\bibitem[Durrer(2010)]{durrer10}
R.~Durrer.
\newblock Gravitational waves from cosmological phase transitions.
\newblock \emph{Journal of Physics: Conference Series}, 222:\penalty0 012021,
  2010.

\bibitem[Dutta et~al.(2001)Dutta, Feldbaum, Walz-Flannigan, Guest, and
  Raithel]{dutta01}
S.~K. Dutta, D.~Feldbaum, A.~Walz-Flannigan, J.~R. Guest, and G.~Raithel.
\newblock High-angular-momentum states in cold rydberg gases.
\newblock \emph{Phys. Rev. Lett.}, 86:\penalty0 3993--3996, 2001.

\bibitem[Einstein(1916)]{einstein16}
A.~Einstein.
\newblock Naherungsweise integration der feldgleichungen der gravitation.
\newblock \emph{Sitzungsber. K. Preuss. Akad. Wiss.}, 1:\penalty0 688, 1916.

\bibitem[Einstein(1918)]{einstein18}
A.~Einstein.
\newblock Uber gravitationswellen.
\newblock \emph{Sitzungsber. K. Preuss. Akad. Wiss.}, 1:\penalty0 154, 1918.

\bibitem[\emph{et al.}(2016{\natexlab{a}})]{abbot16}
B.~P.~Abbott \emph{et al.}
\newblock Observation of gravitational waves from a binary black hole merger.
\newblock \emph{Phys. Rev. Lett.}, 116:\penalty0 061102, 2016{\natexlab{a}}.

\bibitem[\emph{et al.}(2016{\natexlab{b}})]{abbot17}
B.~P.~Abbott \emph{et al.}
\newblock Gw151226: Observation of gravitational waves from a 22-solar-mass
  binary black hole coalescence.
\newblock \emph{Phys. Rev. Lett.}, 116:\penalty0 241103, 2016{\natexlab{b}}.

\bibitem[\emph{et al.}(2017)]{abbot18}
B.~P.~Abbott \emph{et al.}
\newblock Gw170104: Observation of a 50-solar-mass binary black hole
  coalescence at redshift 0.2.
\newblock \emph{Phys. Rev. Lett.}, 118:\penalty0 221101, 2017.

\bibitem[\emph{et al.}(2010)]{kubler10}
H.~Kubler \emph{et al.}
\newblock Coherent excitation of rydberg atoms in micrometre-sized atomic
  vapour cells.
\newblock \emph{Nat. Photon.}, 4:\penalty0 112--116, 2010.

\bibitem[\emph{et al.}(2016{\natexlab{c}})]{lasky16}
Paul D.~Lasky \emph{et al.}
\newblock Gravitational-wave cosmology across 29 decades in frequency.
\newblock \emph{Phys. Rev. X}, 6:\penalty0 011035, 2016{\natexlab{c}}.

\bibitem[Fischer(1994)]{fischert94}
U.~Fischer.
\newblock Transition probabilities for a rydberg atom in the field of a
  gravitational wave.
\newblock \emph{Classical Quantum Gravity}, 11:\penalty0 463, 1994.

\bibitem[Gnedin et~al.(2009)Gnedin, Mihajlov, Ignjatovic, Sakan, Sreckovic,
  Zakharov, Bezuglov, and Klycharev]{gnedin09}
Yu.~N. Gnedin, A.~A. Mihajlov, Lj.~M. Ignjatovic, N.~M. Sakan, V.~A. Sreckovic,
  M.~Yu. Zakharov, N.~N. Bezuglov, and A.~N. Klycharev.
\newblock Rydberg atoms in astrophysics.
\newblock \emph{New Astronomy Reviews}, 53:\penalty0 259--265, 2009.

\bibitem[Griffiths(1995)]{griffiths}
D.~Griffiths.
\newblock \emph{Introduction to Quantum Mechanics}, page 150.
\newblock New Jersey: Pearson Education, 1995.

\bibitem[Grishchuk(2005)]{grishchuk05}
L.~P. Grishchuk.
\newblock Relic gravitational waves and cosmology.
\newblock \emph{Physics-Uspekhi}, 48:\penalty0 1235, 2005.

\bibitem[Haehnelt(1994)]{haehnelt94}
M.~G. Haehnelt.
\newblock Low-frequency gravitational waves from supermassive black holes.
\newblock \emph{Monthly Notices of the Royal Astronomical Society},
  2692:\penalty0 199--208, 1994.

\bibitem[Hezel et~al.(1992)Hezel, Burkhardt, Ciocca, He, and
  Leventhal]{hezel92}
T.~P. Hezel, C.~E. Burkhardt, M.~Ciocca, L‐W. He, and J.~J. Leventhal.
\newblock Classical view of the properties of rydberg atoms: Application of the
  correspondence principle.
\newblock \emph{American Journal of Physics}, 60:\penalty0 329, 1992.

\bibitem[Hogan(1998)]{hogan98}
C.~J. Hogan.
\newblock Cosmological gravitational wave backgrounds.
\newblock \emph{AIP Conference Proceedings}, 456:\penalty0 79--86, 1998.

\bibitem[Hogan(2000)]{hogan00}
C.~J. Hogan.
\newblock Gravitational waves from mesoscopic dynamics of the extra dimensions.
\newblock \emph{Phys. Rev. Lett.}, 85:\penalty0 2044--2047, 2000.

\bibitem[Johnson et~al.(2008)Johnson, Urban, Henage, Isenhower, Yavuz, Walker,
  and Saffman]{johnson08}
T.~A. Johnson, E.~Urban, T.~Henage, L.~Isenhower, D.~D. Yavuz, T.~G. Walker,
  and M.~Saffman.
\newblock Rabi oscillations between ground and rydberg states with
  dipole-dipole atomic interactions.
\newblock \emph{Phys. Rev. Lett.}, 100:\penalty0 113003, 2008.

\bibitem[Lee et~al.(2012)Lee, Haffner, and Cross]{lee12}
T.~E. Lee, H.~Haffner, and M.~C. Cross.
\newblock Collective quantum jumps of rydberg atoms.
\newblock \emph{Phys. Rev. Lett.}, 108:\penalty0 023602, 2012.

\bibitem[Lukin et~al.(2001)Lukin, Fleischhauer, Cote, Duan, Jaksch, Cirac, and
  Zoller]{lukin01}
M.~D. Lukin, M.~Fleischhauer, R.~Cote, L.~M. Duan, D.~Jaksch, J.~I. Cirac, and
  P.~Zoller.
\newblock Dipole blockade and quantum information processing in mesoscopic
  atomic ensembles.
\newblock \emph{Phys. Rev. Lett.}, 87:\penalty0 037901, 2001.

\bibitem[Oniga and Wang(2016)]{oniga16}
T.~Oniga and Charles H.-T. Wang.
\newblock Quantum gravitational decoherence of light and matter.
\newblock \emph{Phys. Rev. D}, 93:\penalty0 044027, 2016.

\bibitem[Parker and Pimentel(1982)]{parker82}
L.~Parker and L.~O. Pimentel.
\newblock Gravitational perturbation of the hydrogen spectrum.
\newblock \emph{Phys. Rev. D}, 25:\penalty0 3180--3190, 1982.

\bibitem[Pinto(1995)]{pinto95}
F.~Pinto.
\newblock Rydberg atoms as gravitational-wave antennas.
\newblock \emph{Gen. Relat. Gravit.}, 27:\penalty0 9, 1995.

\bibitem[Poisson(1993)]{poisson93}
E.~Poisson.
\newblock Gravitational radiation from a particle in circular orbit around a
  black hole. i. analytical results for the nonrotating case.
\newblock \emph{Phys. Rev. D}, 47:\penalty0 1497--1510, 1993.

\bibitem[Quinones and Varcoe(2016)]{quinones16}
D.~A. Quinones and B.~Varcoe.
\newblock Decoherence in excited atoms by low-energy scattering.
\newblock \emph{Atoms}, 4\penalty0 (28), 2016.

\bibitem[Ryan(1995)]{ryan95}
F.~D. Ryan.
\newblock Gravitational waves from the inspiral of a compact object into a
  massive, axisymmetric body with arbitrary multipole moments.
\newblock \emph{Phys. Rev. D}, 52:\penalty0 5707--5718, 1995.

\bibitem[Ryan(1997)]{ryan97}
F.~D. Ryan.
\newblock Accuracy of estimating the multipole moments of a massive body from
  the gravitational waves of a binary inspiral.
\newblock \emph{Phys. Rev. D}, 56:\penalty0 1845--1855, 1997.

\bibitem[Schwarzschild(1916)]{schwarzschild16}
K.~Schwarzschild.
\newblock On the gravitational field of a mass point according to einstein’s
  theory.
\newblock \emph{Sitzungsber. Preuss. Akad. Wiss. Berlin (Math.Phys.)},
  7:\penalty0 189--196, 1916.

\bibitem[Singer et~al.(2004)Singer, Reetz-Lamour, Amthor, Marcassa, and
  Weidemuller]{singer04}
K.~Singer, M.~Reetz-Lamour, T.~Amthor, L.~G. Marcassa, and M.~Weidemuller.
\newblock Suppression of excitation and spectral broadening induced by
  interactions in a cold gas of rydberg atoms.
\newblock \emph{Phys. Rev. Lett.}, 93:\penalty0 163001, 2004.

\bibitem[Taylor(1994)]{taylor94}
J.~H. Taylor.
\newblock Binary pulsars and relativistic gravity.
\newblock \emph{Rev. Mod. Phys.}, 66:\penalty0 711--719, 1994.

\bibitem[Turner(1997)]{turner97}
M.~S. Turner.
\newblock Detectability of inflation-produced gravitational waves.
\newblock \emph{Phys. Rev. D}, 55:\penalty0 R435--R439, 1997.

\bibitem[Varcoe et~al.(2000)Varcoe, Brattke, Weidinger, and Walther]{varcoe00}
B.~T.~H. Varcoe, S.~Brattke, M.~Weidinger, and H.~Walther.
\newblock Preparing pure photon number states of the radiation field.
\newblock \emph{Nature}, 403:\penalty0 743--746, 2000.

\bibitem[Walther et~al.(2006)Walther, Varcoe, Englert, and Becker]{walther06}
H.~Walther, B.~T.~H. Varcoe, B.-G. Englert, and T.~Becker.
\newblock Cavity quantum electrodynamics.
\newblock \emph{Rep. Prog. Phys.}, 69:\penalty0 1325, 2006.

\bibitem[Wang et~al.(2016)Wang, Zhang, and Chen]{wang16}
D.~G. Wang, Y.~Zhang, and J.-W. Chen.
\newblock Vacuum and gravitons of relic gravitational waves and the
  regularization of the spectrum and energy- momentum tensor.
\newblock \emph{Phys. Rev. D}, 94:\penalty0 044033, 2016.

\bibitem[Zhao et~al.(2007)Zhao, Liu, and Li]{zhao07}
Z.-H. Zhao, Y.-X. Liu, and X.-G. Li.
\newblock Gravitational corrections to energy-levels of a hydrogen atom.
\newblock \emph{Communications in Theoretical Physics}, 47:\penalty0 4, 2007.

\end{thebibliography}

\end{document}